\documentclass[twocolumn,prX,showpacs]{revtex4}

\usepackage{amssymb,amsmath}
\usepackage{openbib}
\usepackage{amsthm}
\usepackage{epsfig}
\usepackage{subeqnarray}
\usepackage{graphicx}

\newcommand{\ket}[1]{{|#1\rangle}}
\newcommand{\bra}[1]{{\langle#1|}}

\newcommand{\eq}[1]{{eq.(\ref{#1})}}
\begin{document}

%%%%%%%%%%%%%%%%%%%%%%%%%%%%%%%%%
\title{Time Averaged Quantum Dynamics and the Validity of the Effective Hamiltonian Model}
%%%%%%%%%%%%%%%%%%%%%%%%%%%%%%%%%
%%%%%%%%%%%%%%%%%%%%%%%%%%%%%%%%%
\author{Omar Gamel}\email{ogamel@physics.utoronto.ca}\author{Daniel F. V. James}\email{dfvj@physics.utoronto.ca}
\address{Department of Physics, University of Toronto, 60 St. George St., Toronto, Ontario, Canada M5S 1A7.}

%%%%%%%%%%%%%%%%%%%%%%%%%%%%%%%%%
%%%%%%%%%%%%%%%%%%%%%%%%%%%%%%%%%

%%%%%%%%%%%%%%%%%%%%%%%%%%%%%%%%%
%%%%%%%%%%%%%%%%%%%%%%%%%%%%%%%%%

%\shortauthor{Gamel and James}

%\begin{abstract}%%%%%%%%%%%%%%%%%%%%%%%%%
%A general approach to coarse-grained dynamics of closed quantum systems is developed by finding evolution equations of the time-averaged density matrix. The evolution follows a Kraus representation, demonstrating that the time averaging procedure is equivalent to having an environment coupled to our system.
%\end{abstract}%%%%%%%%%%%%%%%%%%%%%%%%%%
%%%%%%%%%%%%%%%%%%%%%%%%%%%%%%%%%

\begin{abstract}
We develop a technique for finding the dynamical evolution in time of an averaged density matrix. The result is an equation of evolution that includes an Effective Hamiltonian, as well as decoherence terms in Lindblad form. Applying the general equation to harmonic Hamiltonians, we confirm a previous formula for the Effective Hamiltonian together with a new decoherence term which should in general be included, and whose vanishing provides the criteria for validity of the Effective Hamiltonian approach. Finally, we apply the theory to examples of the AC Stark Shift and Three-Level Raman Transitions, recovering a new decoherence effect in the latter.
\end{abstract}

\pacs{03.65.Aa, 03.65.Ca}

\maketitle

%%%%%%%%%%%%%%%%%%%%%%%%%%%%%%%%%%
\section{Introduction}
\setcounter{equation}{0}

%\begin{figure}[th!]
%\center{ \epsfig{figure=Fig1.pdf,width=35mm}}
%\caption{A two level system with a detuned interaction.}
%\label{geofig}
%\end{figure}

The density matrix is the mathematical object that carries all measurable statistical information about a quantum system, and therefore, completely characterizes the state - whether pure or mixed. However, in reality, the physical perception of any quantum system takes place over a finite time interval, rather than instantaneously. Thus understanding the behaviour and evolution of the density matrix convolved with some time averaging function is essential for a complete understanding of quantum dynamics, as observed in any realistic circumstances. 

Moreover, often in practical applications, such as light-matter interaction systems, the Hamiltonian has two distinct parts - one that oscillates at a high frequency and one at a much lower frequency. If one were observing the quantum systems with a time resolution which is too slow to discern the high frequency effect, one might idly suppose that the high frequency component of the Hamiltonian should not play much of a role. And yet effects such as the AC Stark Shift or the Lamb Shift, can be ascribed to just such high frequency terms in the Hamiltonian. The purpose of this paper is to formalize this idea, and find the ``Effective Hamiltonian" that has the effect of this high frequency component included, and examine the validity of this approach in detail.

The paper is organized as follows. In section II, we derive a general formula for the evolution of the time averaged density matrix, finding the Effective Hamiltonian and decoherence terms. Applying this theory to the class of Hamiltonians with harmonic disturbances, we derive an equation of evolution in the form of Lindblad's open system dynamics. The formula for the Effective Hamiltonian found confirms previous results \cite{JamesJerke}, and new decoherence terms are found to result from the averaging process. 

Finally, we test this theory on known physical systems, the AC Stark Shift, and the Three-Level Raman Transitions, finding a new decoherence effect in the latter, which is potentially realizable through experiment. 

Similar work was completed by the author in the context of Effective Hamiltonian Theory \cite{JamesJerke}, and by other authors such us Cohen-Tannoudji \cite{CohenT}, and Shore \cite{Shore}. Effective Hamiltonians were also used for various systems, such as ion traps and cavities, by James \cite{James2}, Plenio et al. \cite{Plenio}, M$\phi$lmer and S$\phi$renson \cite{MolmerSorenson}.

\section{Evolution Equation of Average Density Matrix}
\subsection{The Unitary Evolution}

We start by defining the time-average of an operator $\overline{\hat{\mathcal{O}}}(t)$ as 
\begin{equation}
\overline{\hat{\mathcal{O}}}(t)\equiv\int_{-\infty}^{\infty}f(t-t')\hat{\mathcal{O}}(t')dt',
\end{equation}
where $f(t)$ is real-valued, and has unit area. In particular, we are interested in the averaged density operator $\overline{\rho}(t)$. The positivity, unit trace, and hermiticity of $\rho(t)$ imply the same properties for $\overline{\rho}(t)$, meaning it too is a density matrix. This averaged density operator $\overline{\rho}(t)$ can be interpreted in two ways: one as a purely theoretical construct which we calculate as a mathematical tool to discard high frequencies, and the other as physically representing the "perceived" density matrix through any physical interactions with an apparatus that take a finite nonzero time, with the averaging kernel $f(t-t')$ representing the strength of this interaction in the time window involved.

By assumption, $\rho(t)$ represents the evolution of a closed system, and therefore is given by the unitary evolution:
\begin{equation}
\rho(t) = \hat{U}(t,t_0)\rho(t_0) \hat{U}(t,t_0)^{\dag}
\label{unievo}
\end{equation}
where $\hat{U}(t,t_0)$ is the familiar time-ordered evolution operator which satisfies the Schr\"{o}dinger equation, viz., 
%Applying the averaging procedure directly to this expression is somewhat intractable, so in the next section we resort to a workaround employing generalized Bloch vectors. The purpose is to obtain an equation similar to \eq{unievo} involving only $\overline{\rho}(t)$.
%Now, $\hat{U}(t,t_0)$ satisfies the following Schr\"{o}dinger equation
%
\begin{equation}
i\hbar\frac{\partial\hat{U}(t,t_0)}{\partial t} = \hat{H}(t)\hat{U}(t,t_0)
\label{schrodinger}
\end{equation}
%%%%%%%%
%
In order to calculate a series expansion for $\hat{U}(t,t_0)$, we adopt the standard approach of replacing $\hat{H}(t)$ by $\lambda\hat{H}(t)$, where $\lambda$ is a dimensionless expansion parameter. One can imagine $\lambda$ gradually being increased from 0 to 1, representing the Hamiltonian being ``gradually turned on". When $\lambda = 1$ we have the standard solution in \eq{unievo}. This suggests that we can write $\hat{U}(t,t_0)$ as a series in powers of $\lambda$, that is
\begin{equation}
\hat{U}(t,t_0) \equiv \sum^{\infty}_{n=0} \lambda^n\hat{U}_n(t,t_0).
\label{Ulambda}
\end{equation}
Substituting \eq{Ulambda} into \eq{schrodinger} and matching the coefficients of like powers of $\lambda$, we obtain a recursion relation for $\hat{U}_n(t,t_0)$:
\begin{align}
\frac{\partial\hat{U}_0(t,t_0)}{\partial t} &= 0  \nonumber\\
i\hbar\frac{\partial\hat{U}_n(t,t_0)}{\partial t} &= \hat{H}(t)\hat{U}_{n-1}(t,t_0)  \hspace{30pt}     n \ge 1.
\label{Urecursion} 
\end{align}
Given that for $\lambda \rightarrow 0$ we must have $\hat{U}(t,t_0) = \hat{I}$ (the identity operator), then $\hat{U}_0(t,t_0) = \hat{I}$. 

Substituting recursively into \eq{Urecursion} one obtains the standard expansion for the time evolution operator. On substituting from \eq{Ulambda} into \eq{unievo}, one finds an equation of evolution for $\overline{\rho}(t)$ in terms of $U_n$ and powers of $\lambda$. Suppressing for the moment the explicit dependence on $t$ and $t_0$, and denoting $\rho_0 \equiv \rho(t_0)$, we have
\begin{align}
\overline{\rho} &= \sum^{\infty}_{m,n}\lambda^{m+n}\overline{\hat{U}_m\rho_0 \hat{U}_n^{\dag}} \nonumber\\
	&= \sum^{\infty}_k \lambda^k \Big(\sum_{j=0}^{k} \overline{\hat{U}_{k-j} \rho_0 \hat{U}_j^{\dag}} \Big) \nonumber\\
	&\equiv \sum^{\infty}_k \lambda^k \mathcal{E}_k [\rho_0]\nonumber\\
	&\equiv \mathcal{E} [\rho_0],
	\label{edef}
\end{align}
where $\mathcal{E}_k [\rho_0]$ are linear maps acting on $\rho_0$, and defined as the bracketed term in the second line. $\mathcal{E}[\rho_0]$ is the linear map that gives $\overline{\rho}$. Note that from the definition, $\mathcal{E}_0 = \mathcal{I}$, the identity linear map.
\subsection{Inverse of $\mathcal{E}$}
To proceed, we would like to invert this transformation to find $\mathcal{F} [\overline{\rho}]  = \mathcal{E}^{-1} [\overline{\rho}] $ that operates on $\overline{\rho}$ to give $\rho_0$. That is
\begin{align}
\rho_0 &  \equiv \mathcal{F} [\overline{\rho}]\nonumber\\ 
	   & \equiv \sum_k \lambda^k \mathcal{F}_k [\overline{\rho}].
	   \label{fdef}
\end{align}
%
%Then we find the $\mathcal{F}_k$ in terms of the $\mathcal{E}_k$, which are known in terms of integrals of $H(t)$. 
Na\"{i}vely one might invert the time evolution using the unitarity of $\hat{U}(t,t_0)$, i.e. by swapping $t$ and $t_0$ one should obtain the inverse. However is this approach is invalidated by the time averaging, since unitarity no longer holds. Instead, we use the fact that $\mathcal{F}$ and $\mathcal{E}$ together give the identity transformation (i.e. $\mathcal{F} \big[\mathcal{E}[\rho]\big] \equiv \mathcal{I}[\rho]$). Thus, postulating that $\mathcal{F}$ may be expanded in powers of $\lambda$, and comparing coefficients for powers of the latter, we find:
\begin{equation}
\sum^{\infty}_{m,n=0} \lambda^{m+n} \mathcal{F}_m \big[\mathcal{E}_n[\rho]\big] = \lambda^0  \mathcal{I}[\rho], 
\end{equation}
which implies
\begin{equation}
\sum^{\infty}_{k=0} \lambda^k \Big( \sum_{j=0}^k \mathcal{F}_j \big[\mathcal{E}_{k-j}[\rho]\big] \Big) = \lambda^0  \mathcal{I}[\rho] .
\end{equation}
Comparing coefficients of powers $\lambda$, we find the first few terms in the expansion of $\mathcal{F}$ as follows:
\begin{align}
\mathcal{F}_0 &= \mathcal{E}_0 = \mathcal{I}\nonumber\\
\mathcal{F}_1 &= -\mathcal{E}_1\nonumber\\  
\mathcal{F}_2 &= - \mathcal{E}_2 + \mathcal{E}_1 \big[\mathcal{E}_1\big]. 
\end{align}
The recursion relation for higher order terms is provided in the appendix.

\subsection{Approximation up to Second Order}

Next, we find the effective evolution equation of $\overline{\rho}$ in terms of $\lambda$, so we can examine the lowest order terms and see what they tell us about the Effective Hamiltonian. 
%That is, we find an equation that relates the time derivative $\frac{\partial\overline{\rho}(t)}{\partial t}$ to the matrix $\overline{\rho}(t)$ at this time - this is a Markovian Approximation.
%
Differentiating \eq{edef} with respect to time (denoted by a dot), and substituting from \eq{fdef} we have 
\begin{align}
%\overline{\rho}(t) & = \mathcal{E} [\rho(t_0)] \equiv \sum^{\infty}_n \lambda^n \mathcal{E}_n[\rho(t_0)]\nonumber\\
% \equiv  \sum^{\infty}_n \lambda^n i\hbar\mathcal{\dot{E}}_n[\rho(t_0)]
i\hbar\frac{\partial\overline{\rho}(t)}{\partial t} & = i\hbar\mathcal{\dot{E}} [\rho(t_0)] = i\hbar\mathcal{\dot{E}} \big[ \mathcal{F}[ \overline{\rho}(t)]\big] \nonumber\\
%\equiv \sum^{\infty}_{n,m} \lambda^{n+m} i\hbar\mathcal{\dot{E}}_n \big[ \mathcal{F}_m[ \overline{\rho}(t)]\big]  \nonumber\\
& = \sum^{\infty}_k \lambda^k \Big\{ \sum_{j=0}^k i\hbar\mathcal{\dot{E}}_k \big[ \mathcal{F}_{k-j}[ \overline{\rho}(t)]\big] \Big\}. 
 \label{drhodt}
\end{align}
This can be written as
\begin{equation}
i\hbar\frac{\partial\overline{\rho}(t)}{\partial t} = \sum^{\infty}_k \lambda^k \mathcal{L}_k[ \overline{\rho}(t)],
\label{ldef}
\end{equation} 
% 
%where the map $\mathcal{L}_k$ is defined by the expression in the large curly braces in \eq{drhodt}. To do this, we find the relevant 
Evaluating $\mathcal{E}_k$ and $\mathcal{F}_k$ explicitly, we find, up to second order, 
% (done in the appendix), and use them to calculate the $\mathcal{L}_k$ terms up to desired order. We find, up to second order,
%
%L values
\begin{align}
% L_0
\mathcal{L}_0[\rho] &= i\hbar\dot{\mathcal{E}}_0\big[\mathcal{F}_0[\rho]\big] = 0,\label{l0}\\
\nonumber\\
%
% L_1
\mathcal{L}_1[\rho] &= i\hbar\dot{\mathcal{E}}_1\big[\mathcal{F}_0[\rho]\big] + i\hbar\dot{\mathcal{E}}_0\big[\mathcal{F}_1[\rho]\big] \nonumber\\
&=\overline{\hat{H}}\rho - \rho\overline{\hat{H}},\label{l1}\\
\nonumber\\
\mathcal{L}_2[\rho] &=\overline{\hat{H}\hat{U}_1}\rho - \overline{\hat{H}}\,\overline{\hat{U}_1}\rho  + \overline{\hat{H}\rho\hat{U}_1^{\dag}} - \overline{\hat{H}}\rho\overline{\hat{U}_1^{\dag}}    \nonumber\\ 
& - \rho\overline{\hat{U}_1^{\dag}\hat{H}} + \rho\overline{\hat{U}_1^{\dag}}\,\overline{\hat{H}} - \overline{\hat{U}_1\rho\hat{H}} + \overline{\hat{U}_1}\rho\overline{\hat{H}}\label{l2}.
\end{align}
Eq. (\ref{ldef}) together with eqs. (\ref{l0}) - (\ref{l2}) form the principal result of this section. Higher orders are included in the appendix. The $\mathcal{L}_1$ term is just the Heisenberg equation using the averaged Hamiltonian. Note that the total expression for $\mathcal{L}_k$ is anti-Hermitian, as we would expect, since the LHS of \eq{ldef} is a pure imaginary multiplied by a Hermitian operator, making it anti-Hermitian. Also, the familiar pattern of ``average of the product minus product of the averages", in a manner reminiscent of the definition of covariances, occurs repeatedly.
% Compare with old results
\subsection{Comparison with Previous Results}

Previous results due ref. \cite{JamesJerke}, show that to first order, the Effective Hamiltonian is given by
\begin{equation}
\hat{H} _{eff} = \overline{\hat{H}} + \frac{1}{2}(\hat{A} + \hat{A}^{\dag}),
\end{equation}
where
\begin{align}
\hat{A} &= \overline{\hat{H}\hat{U}_1} - \overline{\hat{H}}\,\overline{\hat{U}_1},\\
\hat{A}^{\dag} &= - \overline{\hat{U}_1\hat{H}} + \overline{\hat{U}_1}\,\overline{\hat{H}}.
\end{align}
This holds since $\hat{U}_1^{\dag} = -\hat{U}_1$. Using the von Neumann equation, this implies that to first order:
\begin{equation}
i\hbar\frac{\partial\overline{\rho}}{\partial t} = [\overline{\hat{H}},\overline{\rho}] + \frac{\hat{A}+\hat{A}^{\dag}}{2}\overline{\rho} - \overline{\rho} \frac{\hat{A}+\hat{A}^{\dag}}{2}.
\end{equation}
However, note that the derivation in ref. \cite{JamesJerke} simply discarded anti-Hermitian terms to satisfy the Hermiticity requirement for the Effective Hamiltonian. Our derivation however treats this problem with more rigor, and arrives at the detailed solution. From the previous section up to second order (i.e. including $\mathcal{L}_1$ and $\mathcal{L}_2$), we have
\begin{equation}
i\hbar\frac{\partial\overline{\rho}}{\partial t} = [\overline{\hat{H}},\overline{\rho}] + \hat{A}\overline{\rho} - \overline{\rho} \hat{A}^{\dag} + \mathcal{D}_2[\overline{\rho}] \label{neweq},
\end{equation}
\begin{equation}
i\hbar\frac{\partial\overline{\rho}}{\partial t} = \big[\hat{H}_{eff},\overline{\rho}\big] + \big\{ \frac{\hat{A}-\hat{A}^{\dag}}{2}, \overline{\rho} \big\} + \mathcal{D}_2[\overline{\rho}] \label{neweq},
\end{equation}
where $\mathcal{D}_2[\overline{\rho}]$ are the decoherence terms, i.e. all the terms where $\rho$ is ``sandwiched" in $\mathcal{L}_2$:
\begin{equation}
\mathcal{D}_2[\overline{\rho}] \equiv \overline{\hat{H}\rho\hat{U}_1^{\dag}} - \overline{\hat{H}}\rho\overline{\hat{U}_1^{\dag}} + \overline{\hat{U}_1\rho\hat{H}} + \overline{\hat{U}_1}\rho\overline{\hat{H}}.
\label{decoherenceterms}
\end{equation}

Our results correspond to those of ref. \cite{JamesJerke} if $\hat{A} = \hat{A}^{\dag}$ and the decoherence terms are identically zero. As we shall see later, it turns out the decoherence terms are in general important, and yield an evolution closer to the exact case. Thus eqs. (\ref{ldef}) - (\ref{l2}) are the correct expressions which should be used in general. 
The derivation in this section relied on the Schr\"{o}dinger picture, however it is valid in the Interaction picture as well. In the latter picture however, the interpretation of the density matrix changes somewhat.
\section{Harmonic Time Dependent Hamiltonian}

Following ref. \cite{JamesJerke}, we will now apply our general result to a class of harmonic Hamiltonians of the form:
% picture??
%
\begin{equation}
\hat{H}(t) =\hat{H}_0 + \sum^N_{n=1}\hat{h}_n exp(-i\omega_nt) + \hat{h}_n^{\dag} exp(i\omega_nt),
\label{harmonich}
\end{equation}
where $\hat{H}_0$ is independent of time. This is representative of a wide class of problems, particularly in the interaction picture Hamiltonian. In this case, the first term in the expansion for $\hat{U}_1$ is
\begin{equation}
\hat{U}_1(t) = \frac{(t-t_0)}{i\hbar} \hat{H}_0 + \hat{V}_1(t) - \hat{V}_1(t_0),
\end{equation}
where
\begin{equation}
\hat{V}_1(t) =\sum^N_{n=1}\frac{1}{\hbar\omega_n}\Big\{\hat{h}_n exp(-i\omega_nt) - \hat{h}_n^{\dag} exp(i\omega_nt) \Big\}
\end{equation}
%
% picture??
We assume the averaging kernel $f(\cdot)$ is an ideal low pass filter, and an even function in time. We further assume that the frequencies $\omega_n$ are sufficiently high that they are filtered out, but sufficiently close to each other so that terms oscillating at difference frequencies  ($\omega_n - \omega_m$) will pass the filter. Thus, we can make the following assumptions:
\begin{align}
\overline{\exp{\pm i \omega_{n}t}}&=0 \nonumber\\
\overline{\exp{\pm i (\omega_{n}+\omega_{m})t}}&=0 \nonumber\\
\overline{\exp{\pm i (\omega_{n}-\omega_{m})t}}&=\exp{\pm i (\omega_{n}-\omega_{m})t}. \label{avg0}
\end{align}
This implies that
\begin{align}
\overline{\hat{H}(t)} &= \hat{H}_0, \ \ \ \ \ \ \ \ \ \overline{\hat{V}_1(t)} = 0,\nonumber\\
\overline{\hat{U}_1(t)} &= \frac{(t-t_0)}{i\hbar} \hat{H}_0 - \hat{V}_1(t_0).
%\overline{\hat{H}(t)\hat{U}_1(t)} &=  
\end{align}
Using the Harmonic Hamiltonian \eq{harmonich} along with \eq{avg0} and our initial results in \eq{neweq} and \eq{decoherenceterms}, we obtain the expression
\begin{align}
\label{maineqn}
i\hbar\frac{\partial\overline{\rho}}{\partial t}  &= \Big[ \hat{H}_0,\overline{\rho} \Big] \nonumber\\
&+  \sum^{N}_{n,m}\Big[ \big(\frac{\hat{h}_m^{\dag}\hat{h}_n}{\hbar\omega_n} -\frac{\hat{h}_n\hat{h}_m^{\dag}}{\hbar\omega_m}\big)\overline{\rho} - \overline{\rho}\big(\frac{\hat{h}_m^{\dag}\hat{h}_n}{\hbar\omega_m} -\frac{\hat{h}_n\hat{h}_m^{\dag}}{\hbar\omega_n}\big)\nonumber\\
&+ \frac{\hat{h}_m^{\dag}\overline{\rho}\hat{h}_n + \hat{h}_n\overline{\rho}\hat{h}_m^{\dag}}{\hbar\omega_m} - \frac{\hat{h}_m^{\dag}\overline{\rho}\hat{h}_n + \hat{h}_n\overline{\rho}\hat{h}_m^{\dag}}{\hbar\omega_n}\Big]e^{i(\omega_m-\omega_n)t}\nonumber\\
 &= \Big[ \hat{H}_{eff},\overline{\rho} \Big] \nonumber\\
&+\Big{\{} \sum^{\infty}_{n,m}\frac{1}{\hbar\omega^{-}_{nm}}\{\hat{h}^{\dag}_m,\hat{h}_n\}e^{i(\omega_m-\omega_n)t},\overline{\rho} \Big{\}} \nonumber\\
&- \sum^{N}_{n,m}2\frac{\hat{h}_m^{\dag}\overline{\rho}\hat{h}_n + \hat{h}_n\overline{\rho}\hat{h}_m^{\dag}}{\hbar\omega^{-}_{nm}}e^{i(\omega_m-\omega_n)t}
\end{align}
where
\begin{equation}
\frac{1}{\omega^{\pm}_{nm}} = \frac{1}{2}\Big(\frac{1}{\omega_n} \pm \frac{1}{\omega_m} \Big)
\end{equation}
and
\begin{equation}
\hat{H}_{eff} = \hat{H}_0 + \sum^{N}_{n,m}\frac{1}{\hbar\omega^{+}_{nm}}[\hat{h}^{\dag}_m,\hat{h}_n]e^{i(\omega_m-\omega_n)t}.
\end{equation}
We can then formulate \eq{maineqn} in Lindblad form:
\begin{align}
i\hbar\frac{\partial\overline{\rho}}{\partial t} = &\big[\hat{H}_{eff}, \overline{\rho} \big] \nonumber\\
&+ \sum^{\infty}_{n,m}\frac{1}{\hbar\omega^{-}_{nm}} \Big(\{\hat{L}^{\dag}_m\hat{L}_n, \overline{\rho}\} - 2\hat{L}_n\overline{\rho}\hat{L}_m^{\dag} \nonumber\\
&+ \{\hat{L}_n\hat{L}^{\dag}_m, \overline{\rho}\} - 2\hat{L}^{\dag}_m\overline{\rho}\hat{L}_n \Big),
\label{lindblad}
\end{align}
where
\begin{equation}
\hat{L}_m = \hat{h}_m e^{-i\omega_mt}.
\end{equation}

%We can diagonalize \eq{lindblad} to yield the standard Lindblad form, provided we can diagonalize the matrix whose $(n,m)$ is given by $1/\omega^{-}_{nm}$. 
The Effective Hamiltonian obtained here is exactly the same one in ref. \cite{JamesJerke}. In addition, we have found the expression for the decoherence terms. If we have only one frequency present in the Hamiltonian, the decoherence terms vanish since $1/\omega^{-}_{mm} = 0$, and the Effective Hamiltonian alone perfectly describes the evolution to second order. In more general cases, the decoherence term must be considered.

The appearance of decoherence Lindblad terms above means the average system can be thought of as an open system, even though the underlying ``real system" is closed, and evolves unitarily. The interpretation of these terms in general may be surprising. When one averages quantities, one is of course throwing away information, and thus increasing the entropy of the system. For example,  a low-pass frequency filtering removes information about high-frequency processes. Naturally, this increasing entropy will lead to decoherence terms in the evolution equations for the average.

One point we have neglected so far is the truncation of our series at second order. We can obtain a sufficiency condition for the legitimacy of this approximation simply by considering the ratio of higher order terms of the expansion. We accomplish this by assuming that $\frac{\eta}{\omega_n} << 1$ $\forall n$, where $\eta$ is the largest eigenvalue of $H$. This ensures higher order terms are progressively smaller, and thus may be safely discarded. 

\section{Examples}

In this section, we revisit two examples cited in ref. \cite{JamesJerke}, namely the AC Stark Shift and Three-level Raman Transitions. Note that for both examples, we work in the interaction picture \cite{Townsend}, starting with the interaction Hamiltonian. Inclusion of the decoherence terms given in \eq{lindblad} will illustrate their importance and elucidate the validity of the Effective Hamiltonian model.
%% picture?

\subsection{AC Stark Shift}

Consider a two level atom interacting with an external harmonic force. The Interaction Hamiltonian is given by,
\begin{align}
\hat{H}_{AC}(t)&=\frac{\hbar \Omega}{2}\{\ket{2}\bra{1}\exp(-i\Delta t)+\ket{1}\bra{2}\exp(i\Delta t)\}.
\end{align}
Applying \eq{maineqn} to $\hat{H}_{AC}(t)$, and evaluating $\hat{H}_{eff}$, we find 
\begin{equation}
\hat{H}_{eff}=-\frac{\hbar \Omega^2}{4 \Delta}(\ket{2}\bra{2}-\ket{1}\bra{1}).
\end{equation}
%5
We also find that all decoherence terms vanish, since there is only one harmonic operator, $\hat{h}_1=\frac{\hbar\Omega}{2}\ket{2}\bra{1}$, and therefore $1/\omega^{-}_{11}$ vanishes by definition. So in this case, the Effective Hamiltonian is time independent.
% Schrodinger picture?

To show the Effective Hamiltonian Theory in action, we compare both the exact and time-averaged evolution of the density matrix for a specific numerical example. Recall that we are in the interaction picture, and therefore the density matrix differs from the standard one in the Schr\"{o}dinger picture. The exact evolution is given by
\begin{equation}
i\hbar\frac{\partial \rho(t)}{\partial t} = [\hat{H}_{AC}(t), \rho(t)],
\end{equation}
and time averaged evolution via the Effective Hamiltonian follows
\begin{equation}
i\hbar\frac{\partial \rho(t)}{\partial t} = [\hat{H}_{eff}, \rho(t)].
\end{equation}

We set the ratio $b \equiv \frac{\Omega}{\Delta}$, which represents the strength of the applied AC field, and measure time in units of the characteristic time $\frac{1}{\Delta}$. Then we start with an arbitrary density matrix, and plot the real part of the off diagonals - i.e. the real part of the coherences between the populations in the two levels. We vary b, and plot the result below for b = 0.3.
%
%\begin{figure}[th!]
%\center{ \epsfig{figure=ACb0-1.pdf,width=50mm}}
%\caption{Ratio b = 0.1}
%\label{b01}
%\end{figure}

\begin{figure}[h!]
%\center{ \epsfig{figure=ACb0-1.pdf,width=50mm}}
 \centerline{\includegraphics[trim =10mm 80mm 10mm 80mm, clip, width=\columnwidth]{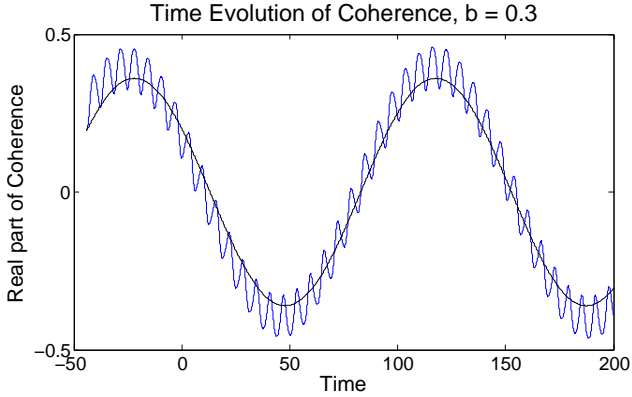}} 
% \centerline{\includegraphics[trim = 1mm 110mm 1mm 110mm, clip, width=\columnwidth]{ACgraph0-3.pdf}} 
\caption{The evolution of the real part of the off diagonal entry for the density matrix, with the interaction strength set at b = 0.3. Both axes dimensionless. }
\label{b01}
\end{figure}

%\begin{figure}[h!]
% \centerline{\includegraphics[trim = 1mm 60mm 1mm 60mm, clip, width=\columnwidth]{ACb0-03.pdf}} 
%%\center{ \epsfig{figure=ACb0-03.pdf,width=50mm}}
%\caption{Ratio b = 0.03}
%\label{b03}
%\end{figure}

%\begin{figure}
%\centering
%\includegraphics{Figure name without .eps extension}
%\caption{Insert caption}
%\end{figure} 

% To find the Schrodinger picture density matrix, we apply the following formula

%\begin{equation}
%\label{i2s}
%\rho_s(t) = e^{-i\hat{H}_0t/\hbar} \rho_I(t) e^{i\hat{H}_0t/\hbar}
%\end{equation}

From the close correspondence graphs, it is clear this method is an effective first order approximation to the frequency and amplitude of oscillation. As expected, the time averaged evolution viz the Effective Hamiltonian resembles the exact evolution without the superimposed high frequency components. 
Note that the relative phase of the exact and averaged evolution just depends on initial conditions, and zero of time. Initial conditions were artificially modified in the above figure so the two graphs are in phase.

\subsection{Three-Level Raman Transitions}

Applying the theory to Raman Transitions, we start with the following Interaction Hamiltonian
\begin{equation}
\hat{H}(t) = \frac{\hbar \Omega_1}{2}\ket{3}\bra{1}e^{-i\omega_1 t}+\frac{\hbar \Omega_2}{2}\ket{3}\bra{2}e^{-i\omega_2 t} + h.a.
\end{equation}
\begin{figure}[h!]
 \centerline{\includegraphics[width=50mm]{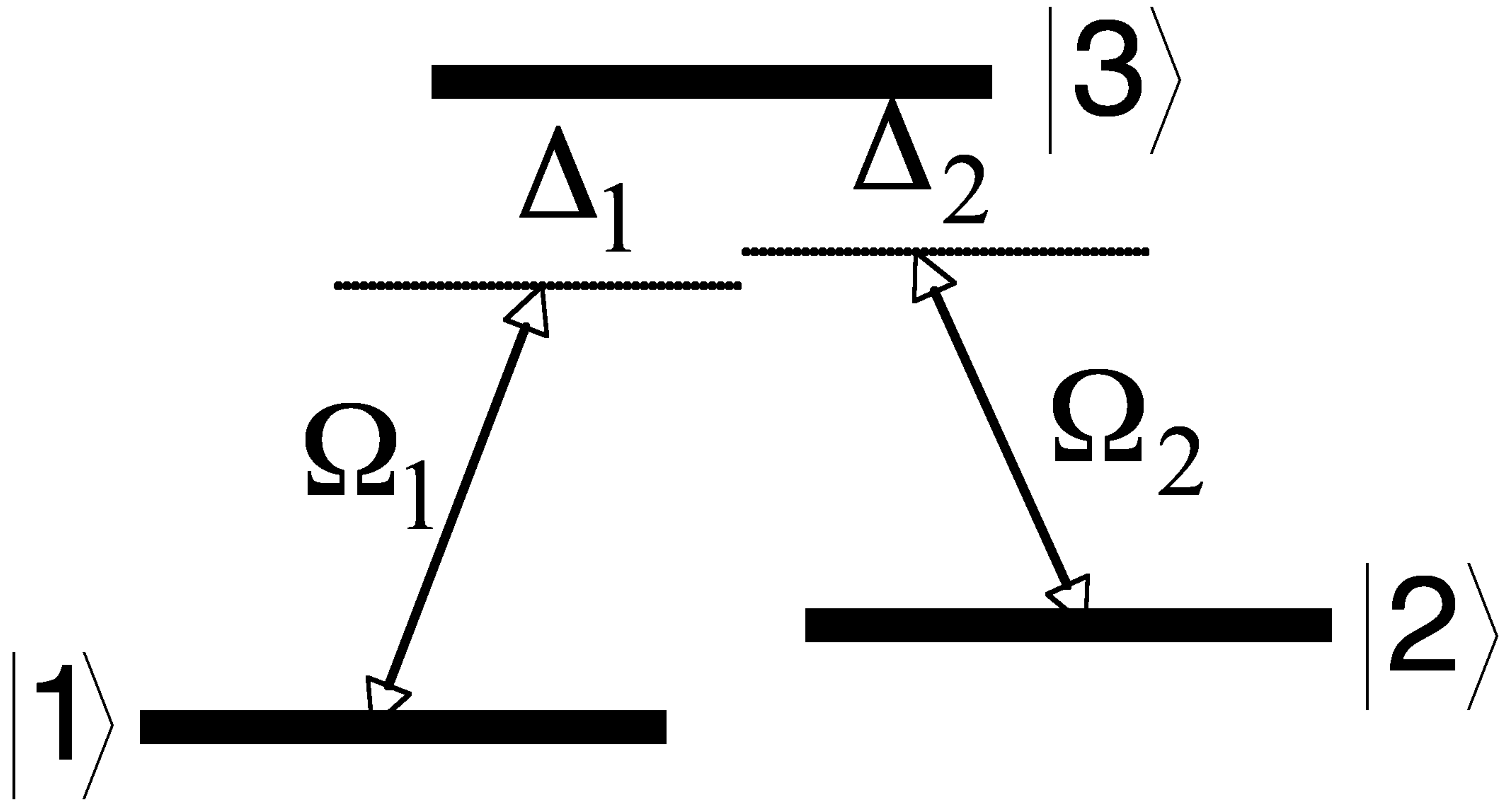}} 
%\center{ \epsfig{figure=ACb0-03.pdf,width=50mm}}
\caption{Illustration of the Three-Level Raman Transitions}
\label{b03}
\end{figure}

Applying our formula \eq{maineqn}, we once again get the same Effective Hamiltonian
\begin{align}
\hat{H}_{eff}= &-\frac{\hbar \Omega_1^2}{4 \omega_1}\big(\ket{3}\bra{3}-\ket{1}\bra{1}\big)-
\frac{\hbar \Omega_2^2}{4 \omega_2}\big(\ket{3}\bra{3}-\ket{2}\bra{2}\big) \nonumber \\
&+\frac{\hbar \Omega_1\Omega_2}{4 \omega^{+}_{12}}
\big(\ket{1}\bra{2} e^{i(\omega_1-\omega_2) t}-\ket{2}\bra{1} e^{-i(\omega_1-\omega_2) t}\big)
\end{align}
The evolution of the density matrix then follows:
%\begin{align}
%i\hbar\frac{\partial\overline{\rho}}{\partial t} &= \big[\hat{H}_{eff}, \overline{\rho} \big] \nonumber\\
%&+ \frac{\hbar\Omega_1\Omega_2}{4\omega^{-}_{12}}\Big[ \big( \{\ket{2}\bra{1},\overline{\rho}\} - 2\ket{3}\bra{1}\overline{\rho}\ket{2}\bra{3} \nonumber\\
%&- 2\ket{2}\bra{3}\overline{\rho}\ket{3}\bra{1} \big)e^{i(\omega_1-\omega_2) t} 
%-\big( \{\ket{1}\bra{2},\overline{\rho}\} \nonumber\\
%&- 2\ket{3}\bra{2}\overline{\rho}\ket{1}\bra{3} - 2\ket{1}\bra{3}\overline{\rho}\ket{3}\bra{2} \big)e^{-i(\omega_1-\omega_2) t}\Big]
%&= \big[\hat{H}_{eff}, \overline{\rho} \big] \nonumber\\
%\end{align}
%
\begin{align}
i\hbar\frac{\partial\overline{\rho}}{\partial t} &= \big[\hat{H}_{eff}, \overline{\rho} \big] + \frac{\hbar\Omega_1\Omega_2}{4\omega^{-}_{12}}\Big[ \big( \{\ket{2}\bra{1},\overline{\rho}\} - 2\overline{\rho}_{12}\ket{3}\bra{3} \nonumber\\
& - 2\overline{\rho}_{33}\ket{2}\bra{1} \big)e^{i(\omega_1-\omega_2) t} -\big( \{\ket{1}\bra{2},\overline{\rho}\} - 2\overline{\rho}_{21}\ket{3}\bra{3}\nonumber\\
& - 2\overline{\rho}_{33}\ket{1}\bra{2} \big)e^{-i(\omega_1-\omega_2) t}\Big],
\label{evol}
\end{align}
where
\begin{equation}
\overline{\rho}_{ij} \equiv \bra{i}\overline{\rho}\ket{j}.
\end{equation}
We already assumed that $\omega_1$ and $\omega_2$ are close in value, which means $\frac{1}{\omega^{-}_{12}}$ will be small compared to $\hat{H}_{eff}$, making the decoherence terms relatively small as expected. Now, we write $\overline{\rho}$ as a generalized Bloch vector,
\begin{align}
\overline{\rho} &= I + r_x X + r_y Y + r_z Z + r_w W \nonumber\\
		&+ r_{xa} X_a + r_{ya} Y_a + r_{xb} X_b + r_{yb}  Y_b
\label{bloch}
\end{align}
where 
\begin{align}
&X = \ket{1}\bra{2} + \ket{2}\bra{1} \ \ \  &Y = -i(\ket{1}\bra{2} - \ket{2}\bra{1})\nonumber\\
&Z = \ket{1}\bra{1} - \ket{2}\bra{2} \ \ \ &W = \frac{1}{\sqrt{3}}(\ket{1}\bra{1} + \ket{2}\bra{2} - 2\ket{3}\bra{3})\nonumber\\
&X_a = \ket{1}\bra{3} + \ket{3}\bra{1} \ \ \  &Y_a = -i(\ket{1}\bra{3} - \ket{2}\bra{3})\nonumber\\
&X_b = \ket{2}\bra{3} + \ket{3}\bra{2} \ \ \  &Y_b = -i(\ket{2}\bra{3} - \ket{3}\bra{2}).
\end{align}
The operators above are the Gell-Mann matrices, and form a basis for SU(3). The operators $X, Y,$ and $Z$ are the standard Pauli operators for the $\{\ket{1},\ket{2}\}$ subsystem. Their analogues for the $\{\ket{1},\ket{3}\}$ and $\{\ket{2},\ket{3}\}$ subsystems are labeled by the subscripts $a$ and $b$ respectively, and will soon be discarded. Substituting \eq{bloch} in \eq{evol}, it simplifies to the following equation:
\begin{align}
\frac{\partial\overline{\rho}}{\partial t} &= \frac{\Omega_1\Omega_2}{2\omega^{+}_{12}}\Big[Z(r_x \sin{\theta} + r_y \sin{\theta}) - r_z\big(Y\cos{\theta} + X\sin{\theta}\big)\Big] \nonumber\\ 
&+ \frac{1}{4}\big(\frac{\Omega_1^2}{\omega_1} - \frac{\Omega_2^2}{\omega_2}\big)\big(r_xY - r_yX\big) \nonumber\\
&-\sqrt{3}\frac{\Omega_1\Omega_2}{2\omega^{-}_{12}}\Big[r_w\big(X\sin{\theta} + Y\cos{\theta}\big) \nonumber\\
&+W\big(r_x\sin{\theta} + r_y\cos{\theta}\big) \Big] + g(...)
\end{align}
where
\begin{equation}
\theta \equiv (\omega_1 - \omega_2)t,
\end{equation}
and $g$ is some linear functional that only depends on $r_{xa}, X_a, r_{ya}, Y_a, r_{xb}, X_b, r_{yb},$ and $Y_b$. This implies that the coherences between level 3 and and the other two levels (i.e. the (1,3), (2,3), (3,1), and (3,2) entries of the density operator) form a closed subsystem, and only affect each other, evolving independently of the rest of the density matrix. Keeping this in mind, we ignore this subsystem entirely, and focus on the evolution of $r_x$, $r_y$, $r_z$, $r_w$, which together form a 4 dimensional analogue of the Bloch vector. This vector follows the evolution of the following dynamical system, viz:
\begin{equation}
\frac{\partial\bold{r}}{\partial t} = A\bold{r} \ \ \ \ \ \ \ \ \ \bold{r}=\left[ \begin{array}{c} r_x\\ r_y\\ r_z  \\ r_w \end{array} \right]
\end{equation}
where
\begin{equation}
A = \left[ \begin{array}{cccc} 0 & -\alpha & -\beta \sin{\theta} & -\gamma \sin{\theta} \\ \alpha & 0 & -\beta \cos{\theta} & -\gamma \cos{\theta} \\ \beta \sin{\theta} & \beta \cos{\theta} & 0 & 0 \\ -\gamma \sin{\theta} & -\gamma \cos{\theta} & 0 & 0 \end{array} \right].
\end{equation}
The constants, $\alpha$, $\beta$, $\gamma$ are given by
\begin{align}
\alpha &\equiv \frac{1}{4}\big(\frac{\Omega_1^2}{\omega_1} - \frac{\Omega_2^2}{\omega_2}\big) \\
\beta &\equiv \frac{\Omega_1\Omega_2}{2\omega^{+}_{12}}\\
\gamma &\equiv \sqrt{3}\frac{\Omega_1\Omega_2}{2\omega^{-}_{12}}.
\end{align}
As we shall see, we can view $\alpha$ and $\beta$ effectively as driving terms, where $\gamma$ represents reduction in frequency due to decoherence. To simplify the matrix $A$ above, we go to the co-rotating frame, and define:
%
% \frac{\partial\bold{\tilde{r}}}{\partial t} = B\bold{\tilde{r}} \ \ \ \ \
\begin{equation} 
\bold{\tilde{r}}=M_\theta \bold{r},
\end{equation}
where
\begin{equation}
M_\theta = \left[ \begin{array}{cccc} \cos{\theta} & -\sin{\theta} & 0 & 0 \\ \sin{\theta} & \cos{\theta} & 0 & 0 \\ 0 & 0 & 1 & 0 \\ 0 & 0 & 0 & 1 \end{array} \right],
\end{equation}
%
%Then making the following approximation:
%\begin{equation}
%\frac{\partial M_{\theta} \bold{r_\theta}}{\partial t} \approx M_{\theta} \frac{\partial \bold{r_\theta}}{\partial t}
%\end{equation}
%
%We can show this by considering the rate of change of the square of the length of $\bold{r_\theta}$. Dropping the $\theta$ subscript, we find:
%
%\begin{equation}
%\frac{\partial r^2}{\partial t} = \frac{\partial \bold{r^\dag r} }{\partial t} = \frac{\partial \bold{r^\dag}}{\partial t}\bold{r} + \bold{r^\dag}\frac{\partial \bold{r}}{\partial t} = \bold{r^\dag}(B + B^\dag)\bold{r} = -4\gamma r_2 r_w
%\end{equation}
%
% and the matrix $B$ is defined in the appendix. 
We can then define the vector $\bold{d}$ as the first three components of $\bold{\tilde{r}}$, leaving $\tilde{r}_w$ as a scalar. We also define new vectors $\boldsymbol{\gamma}$ and  $\bold{\Omega}$  as follows
%We drop the tilde for convenience ($\tilde{r}_w \rightarrow r_w$). 
%
\begin{equation}
\bold{d} \equiv \left[ \begin{array}{c}\tilde{r}_x\\ \tilde{r}_y\\ \tilde{r}_z  \end{array} \right] \ \ \ \ \ \ \ \ \ \boldsymbol{\gamma} \equiv\left[ \begin{array}{c} 0\\ \gamma\\ 0 \end{array} \right] \ \ \ \ \ \ \ \ \ \bold{\Omega} \equiv \left[ \begin{array}{c} \beta \\ 0\\ \alpha +\omega_1 - \omega_2 \end{array} \right].
\end{equation}
Our equations then reduce to the simply form
\begin{align} 
\label{eqn8}
\bold{\dot{d}} &= \bold{\Omega} \times \bold{d} - \tilde{r}_w \boldsymbol{\gamma} \nonumber\\
\dot{\tilde{r}}_w &= - \boldsymbol{\gamma} \cdot \bold{d}.
\end{align}
This presentation intuitively describes the evolution of the system as the rotation of the bloch vector about $\bold{\Omega}$ through the term given by the cross product (as is already well known with the Rabi oscillation \cite{AllenEberly}) plus a perturbation given by the terms in $\boldsymbol{\gamma}$.
Now to solve this, we define the unit vectors $\bold{e}_{\Omega}$ and $\bold{e}_{\gamma}$ as the orthogonal unit vectors along $\bold{\Omega}$ and $\boldsymbol{\gamma}$ respectively. Together with $\bold{e}_{p} \equiv \bold{e}_{\Omega} \bold{\times}\bold{e}_{\gamma}$, they gave an orthonormal basis for 3d space. Using this basis, we write:
\begin{align}
\bold{d} &= d_{\Omega}\bold{e}_{\Omega} + d_{\gamma}\bold{e}_{\gamma} + d_{p}\bold{e}_{p} \nonumber\\
\bold{\Omega} &= \Omega \bold{e}_{\Omega} \nonumber\\
\boldsymbol{\gamma} &= \gamma \bold{e}_{\gamma}
\end{align}
substituting this ansatz for $\bold{d}$ in \eq{eqn8} and resolving the equation along each unit vector, we get the following system of coupled differential equations: 
\begin{align}
\label{eqn9}
\dot{d}_{\Omega} &= 0 \nonumber\\
\dot{d}_{\gamma} &= -\Omega d_p - \gamma r_w \nonumber \\
\dot{d}_p &= \Omega d_{\gamma} \nonumber\\
\dot{\tilde{r}}_w &= -\gamma d_{\gamma},
\end{align}
from which we get
\begin{equation}
\ddot{d}_{\gamma} = -(\Omega^2 - \gamma^2) d_{\gamma}.
\end{equation}
Solving this simple second order ordinary differential equation, we have
\begin{align}
d_{\gamma}&=R\cos{\omega t} \nonumber\\ 
d_{p} &= R\frac{\Omega}{\omega}\sin{\omega t} - \frac{\gamma}{\Omega}\tilde{r}_{w0},  \nonumber\\
\tilde{r}_w(t) &= - R\frac{\gamma}{\omega}\sin{\omega t} + \tilde{r}_{w0}
\end{align}
where $R$ is some constant oscillation amplitude, and $r_{w0} \equiv r_w(0)$ is a constant, both dependent on the initial conditions. The frequency of oscillation $\omega$ given by
\begin{align}
\omega^2 &= \Omega^2 - \gamma^2 \nonumber\\
		&= (\alpha + \omega_1 - \omega_2)^2 + \beta^2 - \gamma^2
\end{align}
Note that we set the zero of time such that there is no phase angle in the argument of the trigonometric terms. So our final solution is
\begin{align}
\bold{d}(t) &= \bold{e}_{\Omega}d_{\Omega} - \bold{e}_p\frac{\gamma}{\Omega}\tilde{r}_{w0}  +  R \big(\bold{e}_{\gamma} \cos{\omega t} +  \bold{e}_p \frac{\Omega}{\omega} \sin{\omega t}\big)
\end{align}
Firstly note that when $\gamma$ vanishes, we have $\omega = \Omega$, and Bloch vector oscillates at the Rabi frequency. When $\tilde{r}_{w0} = 0$, this oscillation is around the torque vector $\bold{\Omega}$, as in the well known result \cite{AllenEberly}. As $\gamma$ increases, it reduces the frequency of precession of the Bloch vector, and changes the oscillation path from a circular one to an Elliptical one. As $\tilde{r}_{w0}$ increases, the centre of this oscillation shifts in the $\bold{e}_p$ a direction (perpendicular to both $\bold{\Omega}$ and $\boldsymbol{\gamma}$. 
The elliptical nature of the oscillation means the length of the vector is not constant, rather it oscillates at frequency $\omega$. Note that the length squared of the Bloch vector corresponds to the trace of the square of the underlying density matrix - i.e. a measure of its purity/coherence. If we define $l^2$ as
\begin{align}
l^2 &= d_{\Omega}^2 + d_{\gamma}^2 + d_{p}^2.
\end{align}
Taking its time derivative and using \ref{eqn9}, we find
\begin{align}
\frac{d(l^2)}{dt} &= 2(d_{\Omega}\dot{d_{\Omega}} + d_{\gamma}\dot{d_{\gamma}} + d_{p}\dot{d_{p}}) \nonumber\\
&= -2\gamma \tilde{r}_w d_{\gamma} \nonumber\\
&= \frac{\gamma^2}{\omega} R^2 \sin{2\omega t} - 2\gamma R\tilde{r}_{w0}\cos{\omega t}
\end{align}
So in general, the length of the Bloch vector (i.e. the coherence of the effective 2 level system) oscillates at frequency $\omega$, and if $\tilde{r}_{w0}$ vanishes, it oscillates at $2\omega$.

\section{Conclusion}
We have demonstrated a more rigorous derivation of Effective Hamiltonian Theory and its domain of applicability. Additionally, it has been shown that additional terms are created resembling Lindblad evolution for harmonic Hamiltonians - implying that the averaging process introduced a small decoherence factor. Applying this theory to examples such as the AC Stark Shift and Raman Transitions, we find it introduces some minor corrections.

In the future, applying this theory to other known systems and testing its limitations would inform us about its usefulness as a tool. In particular, applying it to systems where the newfound decoherence terms play a larger role would help us better understand their exact role. Additionally, it would be useful to come up with a general interpretation of the Lindblad terms. For example, the averaging process over a given system can be seen as observing an analogous system-reservior pair for some hypothetical reservoir, and then observing the system alone while ``tracing out" the reservoir. 

\section*{Acknowledgement}
This work was funded by the Natural Sciences and Engineering Research Council of Canada (NSERC), and the University of Toronto Physics Department.

%--
%JUNK
%
%\begin{align}
% = \Omega $

%\left[ \begin{array}{cccc} 1 & 2 & 3 & 4 \\ 5 & 6 & 7 & 8 \\ 9 & 10 & 11 & 12 \\ 13 & 14 & 15 & 16 \end{array} \right]

% \nonumber\\
%&=\big[\hat{H}_{eff}, \overline{\rho} \big] + 2i\frac{\hbar\Omega_1\Omega_2}{4\omega^{-}_{12}}\Big[\ket{1}\bra{1}+\ket{2}\bra{2}-2\ket{3}\bra{3}\Big] \times  \nonumber\\
%& \ \ \ \ \Big[\overline{\rho}_x \sin{(\omega_1-\omega_2)t} - \overline{\rho}_y \cos{(\omega_1-\omega_2)t}\Big]
%
%----

\appendix
 \section{Derivation of $\mathcal{F}_k$ and $\mathcal{L}_k$ terms}
% \begin{align}
%\lambda^0 & : \mathcal{F}_0 \big[\mathcal{E}_0\big] = \mathcal{I}\nonumber\\
%\lambda^1 & : \mathcal{F}_0 \big[\mathcal{E}_1\big] +  \mathcal{F}_1 \big[\mathcal{E}_0\big] = 0\nonumber\\
%\lambda^2 & : \mathcal{F}_0 \big[\mathcal{E}_2\big] + \mathcal{F}_1 \big[\mathcal{E}_1\big] + \mathcal{F}_2 \big[\mathcal{E}_0\big]= 0\nonumber\\
%\lambda^n & : \mathcal{F}_0 \big[\mathcal{E}_n\big] + \mathcal{F}_1 \big[\mathcal{E}_{n-1}\big] + \mathcal{F}_2 \big[\mathcal{E}_{n-2}\big]= 0\nonumber\\
%\end{align}

The recursion relation for calculating $\mathcal{F}_n$ is given by:

\begin{equation}
\mathcal{F}_n = -\sum_{j=0}^{n-1} \mathcal{F}_j \big[\mathcal{E}_{n-j}[\rho]\big].
\end{equation}

In the following, will calculate $\mathcal{E}_k$ and $\mathcal{F}_k$ terms for the first few orders.
Using \eq{edef} and we get
%
%% E definitions
\begin{align}
\mathcal{E}_0[\rho] &= \rho \nonumber\\
\mathcal{E}_1[\rho] &= \overline{\hat{U}_1}\rho + \rho\overline{\hat{U}_1^{\dag}}\nonumber\\
\mathcal{E}_2[\rho] &= \overline{\hat{U}_2}\rho + \overline{\hat{U}_1\rho\hat{U}_1^{\dag}} + \rho\overline{\hat{U}_2^{\dag}}\nonumber\\
\mathcal{E}_3[\rho] &= \overline{\hat{U}_3}\rho + \overline{\hat{U}_2\rho\hat{U}_1^{\dag}} + \overline{\hat{U}_1\rho\hat{U}_2^{\dag}} + \rho\overline{\hat{U}_3^{\dag}}.
\end{align}
Applying the differential operator $i\hbar\frac{\partial}{\partial t}$:
%
%% E dot definitions
\begin{align}
i\hbar\dot{\mathcal{E}}_0[\rho] &= 0\nonumber\\
i\hbar\dot{\mathcal{E}}_1[\rho] &= \overline{\hat{H}}\rho - \rho\overline{\hat{H}}\nonumber\\
i\hbar\dot{\mathcal{E}}_2[\rho] &= \overline{\hat{H}\hat{U}_1}\rho + \overline{\hat{H}\rho\hat{U}_1^{\dag}} - \overline{\hat{U}_1\rho\hat{H}} - \rho\overline{\hat{U}_1^{\dag}\hat{H}}\nonumber\\
i\hbar\dot{\mathcal{E}}_3[\rho] &= \overline{\hat{H}\hat{U}_2}\rho + \overline{\hat{H}\hat{U}_1\rho\hat{U}_1^{\dag}} - \overline{\hat{U}_2\rho\hat{H}}  + \overline{\hat{H}\rho\hat{U}_2^{\dag}}\nonumber\\ 
&- \overline{\hat{U}_1\rho\hat{U}_1^{\dag}\hat{H}} - \rho\overline{\hat{U}_2^{\dag}\hat{H}}.
\end{align}
%
%% F definitions
Finally from \eq{fdef}
\begin{align}
\mathcal{F}_0[\rho] &= \rho\\
\mathcal{F}_1[\rho] &= - \overline{\hat{U}_1}\rho - \rho\overline{\hat{U}_1^{\dag}}\nonumber\\
\mathcal{F}_2[\rho] &= - \overline{\hat{U}_2}\rho - \overline{\hat{U}_1\rho\hat{U}_1^{\dag}} - \rho\overline{\hat{U}_2^{\dag}} + \overline{\hat{U}_1}\Big(\overline{\hat{U}_1}\rho + \rho\overline{\hat{U}_1^{\dag}}\Big)\nonumber\\ 
&+ \Big(\overline{\hat{U}_1}\rho + \rho\overline{\hat{U}_1^{\dag}}\Big)\overline{\hat{U}_1^{\dag}}.
\end{align}

The third order term $\mathcal{L}_3$ is found to be
%
% L_3
\begin{align}
\mathcal{L}_3[\rho] &= \overline{\hat{H}\hat{U}_2}\rho  - \overline{\hat{H}}\,\overline{\hat{U}_2}\rho + \overline{\hat{H}\rho\hat{U}_2^{\dag}} - \overline{\hat{H}}\rho\overline{\hat{U}_2^{\dag}}  \nonumber\\
&- \rho\overline{\hat{U}_2^{\dag}\hat{H}} + \rho\overline{\hat{U}_2^{\dag}}\,\overline{\hat{H}}  - \overline{\hat{U}_2\rho\hat{H}} + \overline{\hat{U}_2}\rho\overline{\hat{H}}\nonumber\\
&+ \overline{\hat{H}\hat{U}_1\rho\hat{U}_1^{\dag}} - \overline{\hat{H}\hat{U}_1}\rho\overline{\hat{U}_1^{\dag}} - \overline{\hat{H}\overline{\hat{U}_1}\rho\hat{U}_1^{\dag}} \nonumber\\
& - \overline{\hat{H}}\,\overline{\hat{U}_1\rho\hat{U}_1^{\dag}} + 2\overline{\hat{H}}\,\overline{\hat{U}_1}\rho\overline{\hat{U}_1^{\dag}}\nonumber\\
&- \overline{\hat{U}_1\rho\hat{U}_1^{\dag}\hat{H}} + \overline{\hat{U}_1}\rho\overline{\hat{U}_1^{\dag}\hat{H}} + \overline{\hat{U}_1\rho\overline{\hat{U}_1^{\dag}}\hat{H}} \nonumber\\
&+ \overline{\hat{U}_1\rho\hat{U}_1^{\dag}}\,\overline{\hat{H}} - 2\overline{\hat{U}_1}\rho\overline{\hat{U}_1^{\dag}}\,\overline{\hat{H}}\nonumber\\
&- \overline{\hat{H}\hat{U}_1}\,\overline{\hat{U}_1}\rho + \overline{\hat{H}}\,\overline{\hat{U}_1}\,\overline{\hat{U}_1}\rho - \overline{\hat{H}\rho\overline{\hat{U}_1^{\dag}}\hat{U}_1^{\dag}} + \overline{\hat{H}}\rho\overline{\hat{U}_1^{\dag}}\,\overline{\hat{U}_1^{\dag}}\nonumber\\
&+ \rho\overline{\hat{U}_1^{\dag}}\,\overline{\hat{U}_1^{\dag}\hat{H}} - \rho\overline{\hat{U}_1^{\dag}}\,\overline{\hat{U}_1^{\dag}}\,\overline{\hat{H}} + \overline{\hat{U}_1\overline{\hat{U}_1}\rho\hat{H}} - \overline{\hat{U}_1}\,\overline{\hat{U}_1}\rho\overline{\hat{H}}.
\end{align}

%In the example of the Three-Level Raman Transitions, we find the new matrix $B$ to be
%%
%\begin{align}
%B = \left[ \begin{array}{cccc} 0 & -(\alpha + \omega_1 - \omega_2) & 0 & 0 \\ \alpha + \omega_1 - \omega_2 & 0 & -\beta & -\gamma \\ 0 & \beta & 0 & 0 \\ 0 & -\gamma & 0 & 0 \end{array} \right]
%\end{align}
%%
%This matrix is antisymmetric if $\gamma$ vanishes, as expected when the system is on resonance (i.e. $\omega_1 = \omega_2$). The antisymmetric part of $B$ will preserve the length of the Bloch vector $\bold{\tilde{r}}$ and just lead to its rotation over time, where as the antisymmetric part will change its length over time, meaning the state purity is changing. 

\end{document}